\documentstyle{elsartwb}
\input psfig.sty

\begin{document}

%\hfill INP ..../PH

\begin{frontmatter}
\title{Non-uniform chiral phase in effective chiral quark models}

\author{Mariusz Sadzikowski and Wojciech Broniowski}
\address{H. Niewodnicza\'nski Institute of Nuclear Physics, PL-31342 Krak\'ow, Poland}

\begin{abstract}
We analyze the phase diagram in effective chiral quark models (the
Nambu--Jona-Lasinio model, the $\sigma$-model with quarks) and show that
at the mean-field level a
phase with a periodically-modulated chiral fields separates the usual
phases with broken and restored chiral symmetry. A possible signal
of such a phase is the production of multipion jets travelling
in opposite directions, with individual pions having
momenta of the order of several hundred MeV. This signal can be interpreted
in terms of disoriented chiral condensates.
\end{abstract}


\end{frontmatter}

\vspace{-7mm} PACS: 12.39.Fe, 21.65.+f, 12.38.Mh, 64.70.-p

%\addtolength{\baselineskip}{1.0\baselineskip}

The problem of phase transitions in hadronic matter is one of the most
important issues in strong-interaction physics. It has been studied
theoretically for many years with a variety of methods, with the results
gaining importance at the approach of the operation of RHIC. We now believe
that in the ($T$-$\mu $) diagram, where $T$ denotes the temperature and $\mu 
$ the baryon chemical potential, one can recognize a quite complicated
structure: phases with broken/restored chiral symmetry, confined/deconfined
phases, color superconducting phase (in the case of two light flavors) \cite
{bailin+wilczek}, or color-flavor locked phase (in the case of three light
flavors) \cite{alford}. In certain limits detailed features of phase
transitions are known from first-principle QCD calculations (color
superconducting gaps at asymptotically-large baryon densities \cite{son}),
or general symmetry considerations ({\em e.g.} the universality class of the
chiral phase transition \cite{pisarski} in the massless limit). Lattice
calculations provide information for finite-temperature QCD at zero baryon
density \cite{karsh}. However, these ``exact'' methods are inapplicable to
the case of moderate baryon densities (up to a few times the nuclear saturation density),
where out of necessity we have to rely on models \cite{wf,models}. In fact, a
lot of our experience comes from model calculations, just to mention the
expectation that chiral symmetry is eventually restored as the baryon
density is increased.

In this letter we consider the chiral phase structure for a class of
low-energy chiral models: the Nambu --Jona-Lasinio model \cite{njl} and the
linear $\sigma $-model with quarks \cite{lsigma}. These models have been
used extensively over the past years to describe low-energy properties of
mesons and baryons. They have also been used to describe dense and hot
system of the Fermi gas of quarks \cite{wf}, in particular to study
chiral restoration at high $T$ or $\mu $. What is usually claimed is that at
the mean-field level (synonymous to one-quark-loop or leading-$N_{c}$) the
model has two phases: with broken chiral symmetry, {\em i.e.} large
dynamically-generated quark mass $M$, which appears at low $T$ and $\mu $,
and with restored chiral symmetry, with massless (or almost massless)
quarks, at higher values of $T$ or $\mu $. We argue that the situation is
much more complicated if one allows for more general mean-field solutions of
the model. We show that for a certain range of $T$ and $\mu $ there appears 
{\em a non-uniform phase with broken rotational and isospin symmetry}. In
the $T$-$\mu $ diagram this phase separates the usual chiral-symmetry-broken
phase from the chiral-symmetry-restored phase. The results for the linear $%
\sigma $-model with quarks at $T=0$ were obtained a long time ago by
Kutschera, Kotlorz, and one of us (WB) \cite{kutschera}. Here we generalize
the calculation to finite temperatures, as well as apply it to the
Nambu--Jona-Lasinio model.

The Lagrangian densities of the $\sigma $-model and of the
partially-bosonized Nambu--Jona-Lasinio model have the form $\bar{\psi}%
\left[ i\partial \!\!\!/-g\left( \sigma +i\gamma _{5}{\bf \tau }\cdot {\bf %
\pi }\right) \right] \psi +L_{{\rm m}}\left( \sigma ,{\bf \pi }\right) $,
where $\psi $ denotes the quark fields, $g$ is the quark-meson coupling
constant, and the mesonic part $L_{{\rm m}}\left( \sigma ,{\bf \pi }\right) $
is specific to the model. We consider spatially non-uniform,
time-independent chiral fields: 
\begin{equation}
\sigma (\vec{x})=\frac{M}{g}\cos (\vec{q}\cdot \vec{x}),\;\;\pi _{0}(\vec{x}%
)=\frac{M}{g}\sin (\vec{q}\cdot \vec{x}),\;\;\pi _{\pm }(\vec{x})=0,
\label{cond}
\end{equation}
where $M$ and $\vec{q}$ are parameters, which are determined dynamically.
Such an ansatz has been considered for the first time in Refs. \cite{dautry}
in context of the {\em pion condensation} in nuclear matter. In the case of
vanishing $\vec{q}$ we recover $\sigma =\frac{M}{g}$, $\vec{\pi }=0$. The
Dirac equation with fields (\ref{cond}) can be solved exactly \cite{dautry,kutschera}, 
which facilitates greatly practical \cite{kutschera,broniowski} and
theoretical \cite{broniowski2} applications. The quark spectrum has two
branches: 
\begin{equation}
E_{\pm }(\vec{p})=\sqrt{M^{2}+\vec{p}\,^{2}+\frac{1}{4}\vec{q}\,^{2}\pm 
\sqrt{M^{2}\vec{q}\,^{2}+(\vec{q}\cdot \vec{p})^{2}}},  \label{spec}
\end{equation}
where $\vec{p}$ labels the momentum of the quark. The basic quantity of our
study is the grand thermodynamical potential density $\Omega (T,\mu ;M,\vec{q%
})$. Its global minimum with respect to $M$ and $\vec{q}$ at fixed $T$ and $%
\mu $ determines the ground state of the system subject to conservation of
baryon number. We stress that our calculation is self-consistent, {\em i.e.}
both quark and meson fields are solutions to the equations of motion. At the
mean-field (one-quark-loop) level we have explicitly 

\newpage

\begin{eqnarray}
&&\Omega (T,\mu ;M,\vec{q})=\Omega _{0}(M,\vec{q}) \\
&&-2N_{c}T\sum_{i=\pm }\int \frac{d^{3}p}{(2\pi )^{3}}\ln \left[ \left(
1+e^{(\mu -E_{i}(\vec{p}))/T}\right) \left( 1+e^{-(\mu +E_{i}(\vec{p}%
))/T}\right) \right] ,  \nonumber
\end{eqnarray}
The medium part of $\Omega $ describes the ideal gas of quarks with the
spectrum (\ref{spec}). The vacuum part, $\Omega _{0}$, depends on the model
considered. In the $\sigma $-model 
\begin{equation}
\Omega _{0}^{\sigma }(M,\vec{q})=\frac{m_{\sigma }^{2}F_{\pi }^{2}}{8}\left(
M^{2}/M_{0}^{2}-1\right) ^{2}+\frac{F_{\pi }^{2}}{2}\left(
M^{2}/M_{0}^{2}\right) \vec{q}\,{}^{2},  \label{omsig}
\end{equation}
where the parameter $M_{0}=gF_{\pi }$ is the vacuum value of the constituent quark mass, %
\mbox{$F_{\pi }=93{\rm MeV}$} is the pion decay constant, and $m_{\sigma }$
is the vacuum value of $\sigma $-meson mass, treated as a model parameter.
The first term in (\ref{omsig}) comes from the Mexican Hat potential, while
the second term is the kinetic energy of the meson fields. In the
Nambu--Jona-Lasinio model we have 
\begin{equation}
\Omega _{0}^{{\rm NJL}}(M,\vec{q})=(M^{2}-M_{0}^{2})/(4G)+V(M)+\frac{F_{\pi
}^{2}}{2}\left( M^{2}/M_{0}^{2}\right) \vec{q}\,{}^{2}+{\cal O}(\vec{q}%
\,{}^{4}),  \label{omNJL}
\end{equation}
where $G$ is the four-quark coupling constant, and $V(M)$ is the Dirac sea
contribution to the energy for the case $\vec{q}=0$. Here we adopt the
simple 3-momentum regulator, hence $V(M)=-2N_{c}/\pi ^{2}\int_{0}^{\Lambda
}dp\,p^{2}\left( \sqrt{p^{2}+M^{2}}-\sqrt{p^{2}+M_{0}^{2}}\right) $ , where $%
\Lambda $ is the cut-off parameter. In the calculation presented below we
use the parameters of Ref. \cite{klevansky}: $G=5.01{\rm GeV}^{-2}$, $%
\Lambda =650{\rm MeV}$. Equation (\ref{omNJL}) has been gradient-expanded in
powers of $\vec{q}$. The coefficient of the quadratic term is model independent. Higher-order
terms depend on the value of $\Lambda $. However, their contribution is
small for moderate values of $\vec{q}$ \cite{broniowski2} and we can
safely drop them. With this simplification the two models become very
similar. The same conclusion holds for other variants of the
Nambu--Jona-Lasinio model, which differ by the regularization method. In
fact, the basic difference between the models is the difference of values of $%
\Omega _{0}$ at the minimum and at the point $M=0$, $\vec{q}=0$. The larger
this difference, the more difficult it is to restore chiral symmetry.

\begin{figure}[tbp]
\centerline{%
\psfig{figure=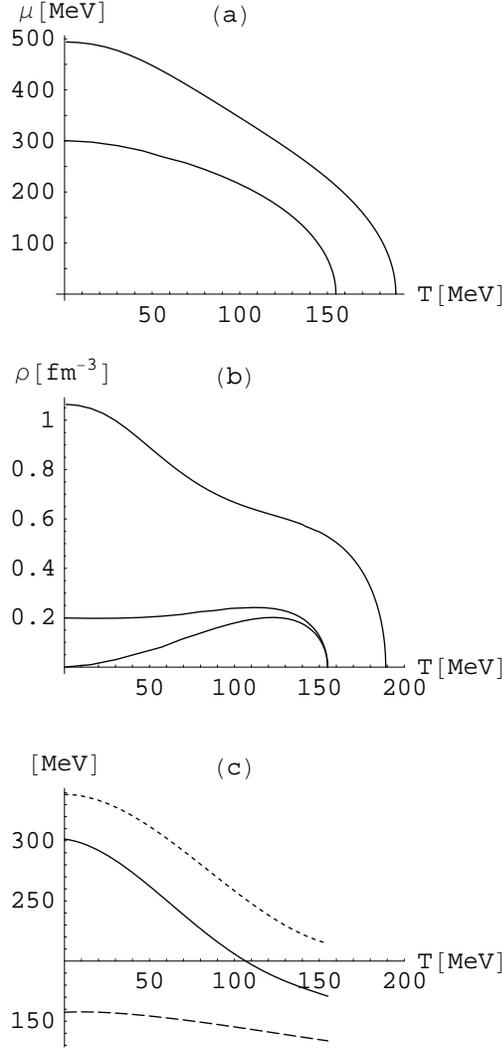,height=15cm,bbllx=8bp,bblly=220bp,bburx=320bp,bbury=700bp,clip=}}
\caption{(a) The $T$-$\protect\mu$ phase diagram for the Nambu--Jona-Lasinio
model with parameters of Ref. \cite{klevansky}. The lower line separates the
uniform and non-uniform chirally-broken phases (first-order phase
transition), and the upper curve separates the non-uniform chirally-broken
phase from the chirally-restored phase (second-order phase transition). (b)
Same as (a) in the $T$--baryon-density space. The ares between the two lower
curves is the region of equillibrium of the uniform and non-uniform
chirally-broken phases. (c) Values of $|\vec{q}|$ and $M$ in the non-uniform
chirally-broken phase at the phase transition (dotted and dashed lines,
respectively), and the value of $M$ in the uniform chirally-broken phase at
the phase transition (solid line).}
\end{figure}

The essence of the dynamics of our system can be understood as follows: as
we increase the chemical potential, it becomes favorable for the system to
develop a non-zero $\vec{q}$. This is because the quark energies on the $%
E_{-}$ branch lower with $\vec{q}$, and the energy gain overcomes the
repulsion of the meson kinetic term in $\Omega _{0}$. Thus, for certain $T$
and $\mu $ the ground state has finite $\vec{q}$. At the further increase of 
$\mu $ the  mass of the quark drops to $0$, and chiral symmetry is
restored. The phase diagram for the Nambu--Jona-Lasinio model with
parameters of Ref. \cite{klevansky} is shown in Fig. 1(a). We can see three
phases: at low $\mu $ and $T$ we have the uniform chirally-broken phase,
with $M>0$ and $\vec{q}$ $=0$, wrapped around it is the phase with $M>0$ and 
$\vec{q}$ $\neq 0$ , and at high $\mu $ or $T$ we find the chirally-restored
phase, with $M=0$. The lower line in Fig. 1(a) describes the{\em \
first-order} phase transition between the uniform and non-uniform chirally
broken phases. The discontinuities in the $M$ and $\vec{q}$ parameters can
be read-off from Fig. 1(c). Correspondingly, in the baryon density --
temperature diagram (Fig. 1(b)) we notice the region of equilibrium of the
two phases (the region between the two lower lines). The upper curve in Fig.
1(a) shows the {\em second-order }phase transition between the non-uniform
chirally broken phase to the chirally-restored phase. At this phase
transition the order parameter $M$ vanishes at non-zero $\vec{q}$.
Interestingly, the character of the transition between the uniform and
non-uniform chiral phases depends on the parameters of the models. In the $%
\sigma $-model for $m_{\sigma }=800$MeV we find a very similar diagram as in
Fig. 1. However for $m_{\sigma }=1200$MeV the phase transition is
second-order. It changes character around $m_{\sigma }=1040$MeV. 

There is always the question of applicability and reliability of a model
calculation like ours. Firstly, we need to have sufficiently high baryon
densities or temperatures to trust the calculation. Physical systems at low
densities consist of nucleons, and we need to dissolve them into quarks for
the model to be applicable. By geometrical arguments we may expect this to
happen at densities of the order of a few nuclear-matter saturation
densities. Note that in Fig. 1 this is the region of the non-uniform
chirally-broken phase, thus we may hope that the model is valid there.
Secondly, the size of the system should be sufficiently large such that our
infinite-size calculation is meaningful. It should be larger than $2\pi /|%
\vec{q}|$ in order to accommodate at least one period of the chiral field.
Using the values of $|\vec{q}|\sim 200-600$ MeV this gives the minimum
size in the range $2-6$ fm, easily accessible in heavy-ion collisions.
In addition, a more elaborate calculation
should incorporate meson fluctuations. At low temperatures pions are easily
excited due to their small mass, and such effects should certainly be
included. Note, however, that inclusion of meson loops in the study of the
quark condensate in Ref. \cite{florkowski} modified, but did not
invalidate the mean-field results. Finally, the effect of the explicit
breaking of chiral symmetry by the finite current quark mass, $m$,
should be incorporated. In studies with $\vec{q}=0$ \cite{wf} this effect
was not quantitatively small. Moreover, with finite $m$ chiral restoration
is no longer a second-order phase transition, but becomes a smooth
cross-over. Another question is whether within the mean-field approximation
the solution with fields (\ref{cond}) is the ground state. What we have verified
is that for a certain range of $T$ and $\mu $ we have a lower-energy state
than the uniform phase. A priori, there may exist yet lower-energy, up-to-now
unknown, mean-field solutions.

We conclude with a discussion of a possible signal of the non-uniform chirally
broken phase. Suppose that sufficiently large domains of the non-uniform
chirally-broken phase are created in the cooling process of the plasma
formed in a heavy-ion collision. Such a domain can be treated as a
classical source of pions, in full analogy to disoriented chiral condensates 
(see e.g. \cite{bjorken}). The coherent decay of such a source would lead to a
characteristic signal in the form of two multi-pion ``jets'', with pion momenta
peaked around $\pm \vec{q}$ (in the rest frame of the source), where $\vec{q}
$ is taken at the phase-transition point, and is typically $200-300{\rm MeV}$%
. Indeed, according to Bjorken's description, pion quanta $\delta \phi
^{a}(x)$ satisfy the equation $\left( \Box + m_{\pi }^{2}\right)
\delta \phi ^{a}(x)=j^{a}(\vec{x})$, where $j^{a}(\vec{x})$ is the source, in our case proportional
to the classical pion field, i.e. to $\sin (\vec{q}\cdot \vec{x})$.
Thus the Fourier transform of $\delta \phi ^{a}$ picks up the components at $%
\pm \vec{q}$. Finite size of the domain spreads out the distribution, but if
the domain has the size of, say, $3$ wavelengths $2\pi /|\vec{q}|$, the half-width
of peaks is only $\sim 20\%$ of $|\vec{q}|$. Of course, larger domains lead
to sharper peaks. If multiple domains contribute, then the signal should be
folded with the momentum distribution of the domain and averaged over
directions. In Eq. (\ref{cond}) we have used the neutral classical pion
field. This case leads to production of neutral pion quanta. However, we can
rotate ansatz (\ref{cond}) in isospin, which leads to a degenerate solution.
Such a configuration will decay into jets of charged pions. In addition, the
quanta of the $\sigma $ field decay subsequently into pairs on neutral as
well as charged pions. A more detailed calculation of the spectrum of pions
requires additional assumptions for the dynamics of the phase transitions
and is left for a further study. Also, the analysis of the formation and
stability of domains is outside of the scope of this paper.

We thank Marek Kutschera, Jacek Dziarmaga, and Wojciech Florkowski for many
useful conversations. One of us (MS) acknowledges the support of the Polish
State Committee for Scientific Research, grant 2P03B 086 14.


\begin{thebibliography}{99}
\bibitem{bailin+wilczek}  D. Bailin and A. Love, {\it Phys. Rep}. {\bf 107}
(1984) 325, M. Alford, K. Rajagopal and F. Wilczek, {\it Phys. Lett}. {\bf %
B422} (1998) 247.

\bibitem{alford}  M. Alford, K. Rajagopal and F. Wilczek, {\it Nucl. Phys}. 
{\bf B537} (1999) 443.

\bibitem{son}  D. T. Son, {\it Phys. Rev}. {\bf D59} (1999) 094019

\bibitem{pisarski}  R. D. Pisarski and F. Wilczek, {\it Phys. Rev}. {\bf D29}
(1984) 338.

\bibitem{karsh}  F. Karsch and E. Laermann, {\it Phys. Rev}. {\bf D50} (1994)
6954.

\bibitem{wf} T. Hatsuda and T. Kunihiro, {\it Phys. Rev. Lett.} {\bf 55}
(1985) 158; V. Bernard, U.-G. Meissner, and I. Zahed, {\it Phys. Rev.} {\bf D36} (1987) 819;
U. Vogel and W. Weise, {\it Prog. Part. Nucl. Phys.} {\bf 27} (1991) 195;
S.P. Klevansky, {\it Rev. Mod. Phys.} {\bf 64} (1992) 649;
M.K. Volkov, {\it Phys. Part. Nucl.} {\bf 24} (1993) 35;
T. Hatsuda and T. Kunihiro, {\it Phys. Rep.} {\bf 247} (1994) 221;
W. Florkowski, {\it Acta Phys. Pol.} {\bf B28} (1997) 2079.

\bibitem{models}   M. A. Halasz et al., {\it Phys. Rev.} {\bf D58} (1998) 096007; 
E.V. Shuryak, {\it Nucl. Phys.} {\bf A642} (1998) 14;
G. W. Carter and D. Diakonov, {\it Phys. Rev.} {\bf D60} (1999) 016004; 
J. Berges and K. Rajagopal, {\it Nucl. Phys.} {\bf B538} (1999) 215; 
R. Rapp, T. Sch\"afer, E.V. Shuryak and M. Velkovsky, SUNY-NTG-99-04 (hep-ph/9904353);
J. Dziarmaga and M. Sadzikowski, {\it Phys. Rev. Lett.} {\bf 82} (1999) 4192.

\bibitem{njl}  Y. Nambu and G. Jona-Lassinio, {\it Phys. Rev}. {\bf 122} (1961)
345; {\bf 124} (1961) 246.

\bibitem{lsigma}  M. Gell - Mann and M. Levy, {\it Nuovo Cimento} {\bf 16}
(1960) 705.
\bibitem{kutschera}  M. Kutschera, W. Broniowski, and A. Kotlorz, 
{\it Phys. Lett.} {\bf B237} (1990) 159; {\it Nucl. Phys}. {\bf A516} (1990) 566.
\bibitem{dautry}  F. Dautry and E.M. Nyman, {\it Nucl. Phys.} {\bf A319}
(1979) 323; D.K. Campbell, R.F. Dashen, and J.T. Manassah, {\it Phys. Rev.}
{\bf D12} (1975) 979, 1010; G. Baym, D.K. Campbell, R.F. Dashen, and J.T. Manassah,
{\it Phys. Lett.} {\bf B58} (1975) 304.
\bibitem{broniowski} W. Broniowski, A. Kotlorz, and M. Kutschera, {\it Acta Physica Pol}. {\bf B22} (1991) 145.
\bibitem{broniowski2} W. Broniowski and M. Kutschera, {\it Phys. Lett.} {\bf B234} (1990) 449; 
{\it Phys. Rev.} {\bf D41} (1990) 3800; {\it Phys. Lett.} {\bf B242} (1990) 133. 
\bibitem{klevansky} T. M. Schwarz, S. P. Klevansky and G. Papp, {\it Phys. Rev.} {\bf C60} (1999) 055205.
\bibitem{florkowski} W. Florkowski and W. Broniowski, {\it Phys. Lett.} {\bf B386} (1996) 62.
\bibitem{bjorken}  J. D. Bjorken, {\it Acta Phys. Pol}. {\bf B28} (1997) 2773.
\end{thebibliography}
\end{document}